\documentclass[reprint, cha]{revtex4-1}

\usepackage{amssymb}
\usepackage{graphicx}
\usepackage{amsmath}
\usepackage{amsfonts}
\usepackage{dsfont}
\usepackage{amssymb}
\usepackage{amsthm}
\usepackage{verbatim}
\usepackage{bm}
\usepackage{natbib}
\usepackage{subfigure}
\usepackage{caption}

\usepackage{color}
\usepackage[utf8]{inputenc}
\usepackage[mathscr]{euscript}
\usepackage{sidecap}

\usepackage{hyphenat}
\usepackage{hyperref}

\usepackage{ulem} 
\usepackage{tikz}

\newcommand{\udensdot}[1]{
    \tikz[baseline=(todotted.base)]{
        \node[inner sep=1pt,outer sep=0pt] (todotted) {#1};
        \draw[densely dotted] (todotted.south west) -- (todotted.south east);
    }
}

\begin{document}

\title{A dynamical systems approach to the discrimination of the modes of operation of cryptographic systems}

\author{Jeaneth Machicao}
 	     \email{machicao@usp.br}
\affiliation{Scientific Computing Group, S\~ao Carlos Institute of Physics, University of S\~{a}o Paulo (USP),  PO Box 369, 13560-970, S\~{a}o Carlos, SP, Brazil - www.scg.ifsc.usp.br}

\author{Jan M. Baetens}
 	     \email{jan.baetens@ugent.be}
\affiliation{KERMIT, Department of Mathematical Modeling, Statistics and Bioinformatics, Ghent University, Coupure links 653, 9000 Ghent, Belgium - www.kermit.ugent.be}

\author{Anderson G. Marco}
\affiliation{ Scientific Computing Group, S\~ao Carlos Institute of Physics, University of S\~{a}o Paulo (USP),  PO Box 369, 13560-970, S\~{a}o Carlos, SP, Brazil - www.scg.ifsc.usp.br}

\author{Bernard De Baets}
 	     \email{bernard.debaets@ugent.be}
\affiliation{ KERMIT, Department of Mathematical Modeling, Statistics and Bioinformatics, Ghent University, Coupure links 653, 9000 Ghent, Belgium - www.kermit.ugent.be}

\author{Odemir M. Bruno}
              \email{bruno@ifsc.usp.br}
\affiliation{  Scientific Computing Group, S\~ao Carlos Institute of Physics, University of S\~{a}o Paulo (USP),  PO Box 369, 13560-970, S\~{a}o Carlos, SP, Brazil - www.scg.ifsc.usp.br }

\date{\today}

\begin{abstract}
Evidence of signatures associated with cryptographic modes of operation is established.
Motivated by some analogies between cryptographic and dynamical systems, in particular with chaos theory, we propose an algorithm based on Lyapunov exponents of discrete dynamical systems to estimate the divergence among ciphertexts as the encryption algorithm is applied iteratively. The results allow to distinguish among six modes of operation, namely ECB, CBC, OFB, CFB, CTR and PCBC using DES, IDEA, TEA and XTEA block ciphers of 64 bits, as well as AES, RC6, Twofish, Seed, Serpent and Camellia block ciphers of 128 bits. Furthermore, the proposed methodology enables a classification of modes of operation of cryptographic systems according to their strength.
\end{abstract}


\maketitle

\section{Introduction}
\label{sec:intro}

The propagation and continuous flow of information are of utter importance for the development of stable economies throughout the world as they are a prerequisite for successful business transactions, short- and long-range communication, and so on~\cite{Shannon}. Often this information has to be encrypted in such a way that it can be safely transferred between the sender and recipient without allowing others to read the information that is present in such an encrypted message~\cite{Stallings}. On the other hand, malicious persons and organizations, but also governmental organizations, are continuously striving to break the key with which messages were encrypted because this might enable them to get those pieces of information that are needed to achieve their criminal, protective, or other goals~\cite{BookCrypto,CodeBook}. It is probably due to the impact of Turing's success in breaking the Enigma that humanity became aware of the importance of cryptography in general, and the vulnerability of ciphers more in particular~\cite{TuringBook}.

Since this major breakthrough, the functioning of the industrial, financial and public sector has become strongly dependent on the advances of cryptography. For instance, while the availability of worldwide networks has enabled rapid dissemination of information, it has also stimulated cryptographic innovations because a significant share of this information may only be available to a few parties. In this manner, technological progress during the last decades has increased the need for secured communication and transactions, information shielding, and so on~\cite{BookCrypto,CodeBook}.

In the last few decades, modern cryptography replaced mechanical schemes with new computing models. This modern focus influenced the classical design of ciphers far beyond the original purpose. Nowadays, there are two class of cryptographic algorithms depending on the key: symmetric and asymmetric. Symmetric encryption algorithms use the same key for both encryption of plaintext and decryption of ciphertext. This class of algorithm is also divided into two categories: stream ciphers and block ciphers. Block ciphers have gained wide popularity since the introduction of the first adopted encryption: The Data Encryption Standard (DES)~\cite{DES}, in the mid-1970s, yet nowadays this cipher is considered prone to brute force attacks. To overcome this shortcoming the International Data Encryption Algorithm (IDEA)~\cite{IDEA} was designed in 1991 to replace DES. Ever since, there has been a pursuit for the development of new algorithms that meet the rising security expectations. In 1997, the National Institute of Standards and Technology (NIST)~\cite{NIST2000} selected the official Advanced Encryption Standard (AES) among many competitors, namely Serpent~\cite{Serpent}, Twofish~\cite{Twofish}, RC6~\cite{RC61998}, Rijndael~\cite{AES2001}, etc.\

To date, block ciphers are the most important elements in many cryptographic systems~\cite{BookCrypto}. A block cipher breaks a message into blocks of elements (bits) and then encrypts one block  (\textit{plaintext}) at a time producing its corresponding output block (\textit{ciphertext}). However, a block cipher by itself allows for the encryption of only one block, such that it is recommended to use a mode of operation in conjunction~\cite{NIST2001}. This mode of operation specifies a mechanism to improve the corresponding block cipher, while encrypting all of the blocks, one by one, as it goes along.

Motivated by the analogies between cryptographic and dynamical systems, on the one hand, and the lack of a means to discriminate between different modes of operation that can be used to encrypt a message with a block cipher using a single key, on the other hand, we demonstrate in this paper how Lyapunov exponents can be relied upon for tackling this problem. More specifically, by contemplating the whole of a cipher, ciphertext and key as an utter discrete dynamical system, \textit{i.e.,} a cellular automaton (CA), and by resorting to the notion of Lyapunov exponents as they have been conceived for such systems~\cite{Baetens2010,Bagnoli92}, we show how these measures can be exploited to identify the mode of operation that was used during the encryption process.

Although the cryptographic process of encrypting and decrypting information does not constitute a dynamical system as such, it has been reported that it is possible to draw parallels between cryptographic and dynamical systems~\cite{Amigo2007,Jakimoski2007,Machicao2012,OTPDynamSyst2012}. Hence, drawing upon such parallels, we have a means to exploit similar tools as the ones that have been conceived in the framework of dynamical systems in order to characterize cryptographic systems. Taking into account that the stability of a dynamical system is generally acknowledged as its main characteristic because it gives insight into its intrinsic nature~\cite{eckmann1985,robinson1998}, it is natural to verify whether the dynamical systems viewpoint of a cryptographic system allows for a similar notion in order to better understand the latter. An exploration of this is further motivated by the fact that several researchers have noticed a close resemblance between a cryptographic system on the one hand, and a chaotic system, on the other hand~\cite{Alvarez2006,Amigo2005,Blackledge2010,Schmitz2008}, and the large number of chaos-based cryptosystems~\cite{Machicao,Anderson}.

Classically, the stability of a dynamical system is assessed by computing its so-called largest Lyapunov exponent that quantifies how it behaves if it is evolved from two different but close initial conditions~\cite{eckmann1985}. Either the corresponding phase space trajectories diverge or converge in which case we refer to the system as unstable or asymptotically stable, respectively, or the system is conservative, which means that the initial separation remains.

As the fields of cryptography and dynamical systems are not yet strongly interwoven, the basic definitions and concepts that relate to those systems and that are of interest within the framework of this paper are presented in Section~\ref{pre}, while the dynamical systems viewpoint on cryptographic systems is presented in Section~\ref{method} together with the proposed method for identifying the underlying mode of operation. Finally, the strengths of the proposed method are illustrated and discussed in Section~\ref{res} by means of computer experiments.

\section{Preliminaries \label{pre}}
In this section we introduce the specificities of both cryptographic and dynamical systems that are indispensable for a clear understanding this paper.

\subsection{Block ciphers and modes of operation}

Classically, an encryption system encloses three major components, namely a cipher, a key, and finally, a ciphertext. The former constitutes a sequence of instructions that must be executed in order to encrypt a given plaintext, which may be envisaged as a sequence of $N$ bits, such that it can be represented as a Boolean vector $\mathbf{P}$ of length $N$. The result of this encryption process using a key $K$, which is a sequence of $k$ bits, is a so-called ciphertext, which may be represented in a similar fashion as a Boolean vector $\mathbf{C}$ of length $N$~\cite{BookCrypto}.

Of course, the real plaintext size varies and is mostly different from the length of the blocks for which a block cipher is designed. Consequently, common ciphers cannot be applied directly for the encryption of arbitrary-length plaintext~\cite{Stallings2010}.
In order to overcome this issue, so-called block ciphers have been designed and implemented. A block cipher slices the plaintext of length $N$ into $b$ blocks of $n$ bits, after which each of these blocks is encrypted/decrypted by a block cipher, denoted as $E_{K}$ and $E^{-1}_{K}$, respectively. Mathematically, the encryption of a plaintext $\mathbf{P}=(P_1,P_2,\ldots,P_b)$ of length $n$ into a ciphertext of the same length can be formalized as  $\mathbf{C}= E(K,\mathbf{P}) = E_{K}(\mathbf{P})$, where $E:\{0,1\}^k\times\{0,1\}^n\to\{0,1\}^n$. If the length $N$ of the plaintext is not a whole multiple of $b$, additional bits are padded to the last block of the plaintext.

These block ciphers encrypt a plaintext in accordance with a well-defined procedure, which is commonly referred to as the mode of operation of a block cipher. A block cipher encrypts one block at a time, and it is the mode of operation that allows a block cipher to encrypt blocks consecutively in a secure way. Most of them use an initialization vector (IV), denoted $\gamma$, which adds randomness to the encryption process~\cite{BookCrypto}. For instance, the counter mode uses a special method for generating counters in order to guarantee that each block in the sequence is different from every other block~\cite{NIST2001}.

From the many block ciphers currently available, we will focus on the well-studied ones, being DES~\cite{DES}~($n=64$, $k=56$), IDEA~\cite{IDEA}, TEA~\cite{TEA} and XTEA~\cite{XTEA} ($n=64$, $k=128$), as well as AES~\cite{AES2001}, RC6~\cite{RC61998}, Twofish~\cite{Twofish}, Serpent~\cite{Serpent}, Seed~\cite{SEED} and Camellia~\cite{Camellia} ($n=128$, $k=128$). The specifications of these block ciphers and the modes of operation are public, so they can be implemented~\cite{NIST2001}.

In 2001 the NIST compiled and recommended five modes of operation---most of them were developed 30 years ago---to be used with a block cipher, namely the Electronic Codebook (ECB), Cipher Block Chaining (CBC), Cipher Feedback (CFB), Output Feedback (OFB), and Counter (CTR) modes~\cite{NIST2001}. In addition to these NIST modes, there are many others, among which the Propagating Cipher-Block Chaining (PCBC)~\cite{BookCrypto} mode will be considered in this paper.

\begin{figure}[b]
  \captionof{table}{Mathematical representation of the CFB, CTR and PCBC modes.}
  \label{modestab}
  \centering\includegraphics[width=1\linewidth,height=1.1in]{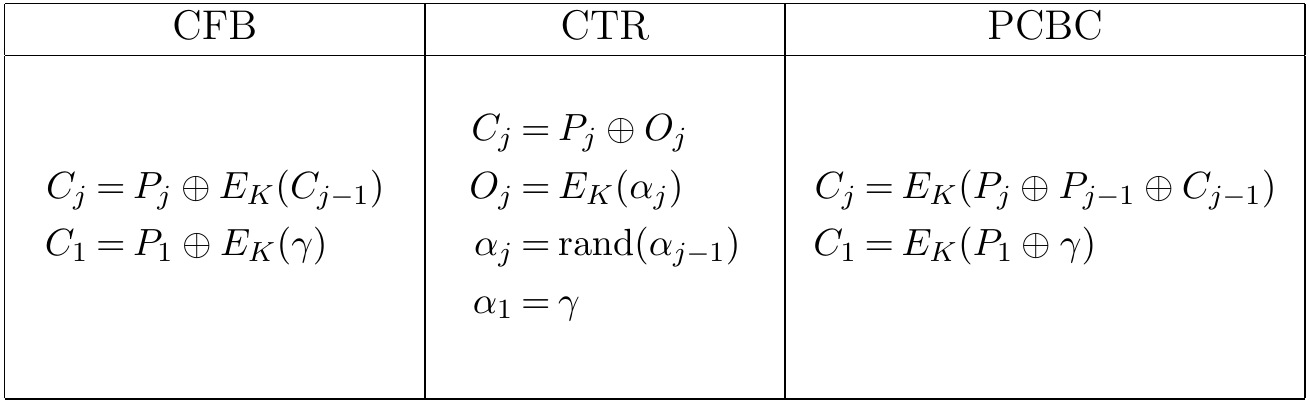}\par
\end{figure}

Mathematically, the  encryption of an arbitrary-length plaintext by means of a block cipher in combination with the ECB mode of operation can be formulated as:

\begin{equation}
C_{j} = \displaystyle  E_{K}(P_{j}) \,,\qquad j=1,2,\ldots,b,
\label{ECBmath}
\end{equation}
where $C_j$ represents the encrypted counterpart of the $j$-th block of plaintext $P_j$. Similarly, the formalism for the CBC mode is given by
\begin{equation}
C_{j}=E_{K}(C_{j-1}\oplus P_{j})\,,\qquad j=1,2,\ldots,b,
\label{CBCMath}
\end{equation}
where $\oplus$ is the mod 2 operator and $C_{0}=\gamma$. For the OFB mode one may write:
\begin{eqnarray}
C_{j}= P_{j}\oplus O_{j-1}\,,\qquad j=1,2,\ldots,b,
\label{OFBmath}
\end{eqnarray}
where $O_{j} = E_{K}(O_{j-1})$ with $O_{0} = E_{K}(\gamma)$ and $\gamma$ is randomly selected from $\left\{0,1\right\}^n$. Table~\ref{modestab} lists the formulas for the other modes of operation that are considered in this paper.

In the remainder, a mode of operation of a block cipher is denoted as $M_{E_K}:\{0,1\}^N\rightarrow \{0,1\}^N$, which maps a given plaintext $\mathbf{P}$ of length $N$ to a corresponding ciphertext $\mathbf{C}$ of equal length, so that we may write $\mathbf{C}=M_{E_K}(\mathbf{P})$. In order to clarify the functioning of a cryptographic system, we show in Fig.~\ref{fig:ex1} a cryptographic system with $b=2$ blocks of $n=3$ bits that uses the CBC mode of operation. We consider an exemplary plaintext $\mathbf{P}=(P_1, P_2)$, for which it holds that $P_1=010$, $P_2=001$ and $\gamma=101$. Note that the outputs generated by $E_K$ are chosen only for illustration, \textit{e.g.,} $C_1= E_K(C_0 \oplus P_1) = E_K(111)=100$. We can see clearly that a single application of the function $M_{E_K}$ consists of $b$ applications of $E_K$ since the latter has to be applied to each of the plaintext blocks $P_j$ in order to construct the ciphertext $\mathbf{C}$.

\newsavebox\A
\begin{lrbox}{\A}
\small
  \begin{minipage}{0.15\textwidth}
  {
     \begin{eqnarray*}
     C_0&=&\gamma=[\udensdot{101}]\nonumber\\
     \mathbf{P}&=&[010,001]
     \\
    \end{eqnarray*}
    }
  \end{minipage}
\end{lrbox}

\newsavebox\B
\begin{lrbox}{\B}
\small
  \begin{minipage}{0.28\textwidth}
  {
     \begin{eqnarray*}
     C_1&=&E_K(\udensdot{101}\oplus 010) = \uline{100}\nonumber\\
     C_2&=&E_K(\uline{100}\hspace{2 dd}\oplus 001) = \textbf{000}\nonumber\\   
     \mathbf{C}&=&M_{E_K}(\mathbf{P})=[100,000]\nonumber
    \end{eqnarray*}
    }
  \end{minipage}
\end{lrbox}

\begin{figure}[!h]
\begin{center}
    \begin{tabular}{c|c}
        \hline
           $t=0$& $t=1$\\
           \hline
           \usebox{\A}&\usebox{\B} \\
          \hline
    \end{tabular}
\caption{Example of the CBC mode of operation with $b=2$ blocks, $n=3$ bits and block cipher $E_{K}$.}
\label{fig:ex1}
\end{center}
\end{figure}

\subsection{Analogies between cryptographic and dynamical systems}
\label{sec:analogies}
Although cryptographic systems, as the ones given by Eqs.~\eqref{ECBmath}--\eqref{OFBmath} and those listed in Table~\ref{modestab}, do not constitute dynamical systems, we can draw some parallels between both types of system, which might enable us to gain a deeper understanding of the former~\cite{Alvarez2006,Amigo2007,Blackledge2010,Jakimoski2007}. More specifically, we may envisage such a cryptographic system  as a one-dimensional CA, which can be represented by means of a triplet $\langle \mathcal{T}, S, \Psi \rangle$. The first element of this triplet refers to a one-dimensional array of `cells' $c_i$, each of which bears one of the states enclosed in the finite set $S=\{0,1\}$, and which are updated at discrete time steps by means of a global transition function $\Psi$. The state of the $i$-th cell in $\mathcal{T}$ at the $t$-th time step will be denoted as $s(c_i,t)$. Essentially, upon putting $\Psi\equiv M_{E_K}$, a mode of operation $M_{E_K}$ may be envisaged as such a global transition function.

Finally, we identify a given plaintext $\mathbf{P}$ with $s(\cdot,0)$ in such a way that the $i$-th bit of the $j$-th block in $\mathbf{P}$ is denoted as $s(c_i^j,0)$. The transition function $M_{E_K}$ may be applied iteratively so that a distinct ciphertext $\mathbf{C}^t$ is evolved at every time step $t$. As such, a cryptographic system can be transformed into a CA, and we may write

\begin{equation}
 s(\cdot, t+1)= M_{E_K}(s(\cdot,t)) \,,
 \label{eq:DynamicSystem}
\end{equation}
 or equivalently, $\mathbf{C}^{t+1} = M_{E_K}(\mathbf{C}^{t})$, where $\mathbf{C}^{1}  = M_{E_K}(\mathbf{P})$.

\newsavebox\D
\begin{lrbox}{\D}
\small
  \begin{minipage}{0.18\textwidth}
  {\setlength\arraycolsep{0.15em}
     \begin{eqnarray*}
     C_0&=&\gamma=[\udensdot{101}]\nonumber\\
     \mathbf{P}&=&[010,001]\\
     \\
     \\
    \end{eqnarray*}
    }
  \end{minipage}
\end{lrbox}

\newsavebox\E
\begin{lrbox}{\E}
\small
  \begin{minipage}{0.28\textwidth}
  {\setlength\arraycolsep{0.15em}
     \begin{eqnarray*}
     C^1_{0}&=&\gamma=[\udensdot{101}]\nonumber\\
     C^1_{1}&=&E_K(\udensdot{101}\oplus 010) = \uline{100}\nonumber\\
     C^1_{2}&=&E_K(\uline{100}\hspace{2 dd}\oplus 001) = \textbf{000}\nonumber\\
     \\
     \mathbf{C^1}&=&M_{E_K}(\mathbf{P})=[100,000]\nonumber
    \end{eqnarray*}
    }
  \end{minipage}
\end{lrbox}

\newsavebox\F
\begin{lrbox}{\F}
\small
  \begin{minipage}{0.3\textwidth}
  {\setlength\arraycolsep{0.15em}
     \begin{eqnarray*}
    \gamma&=& C^2_{0}=C^1_{2}=[\textbf{000}]\nonumber\\
    C^1_{1}&=&E_K(\textbf{000}\oplus 100) = \uline{011}\nonumber\\
    C^1_{2}&=&E_K(\uline{011}\hspace{2 dd}\oplus 000) = 111\nonumber\\
    \\
    \mathbf{C^1}&=&M_{E_K}(\mathbf{P})=[011,111]\nonumber
    \end{eqnarray*}
    }
  \end{minipage}
\end{lrbox}

\begin{figure*}[!ht]
\begin{center}
    \begin{tabular}{c|c|c}
        \hline
           $t=0$& $t=1$ &$t=2$\\
           \hline
           \usebox{\D}&\usebox{\E} & \usebox{\F}\\
           \hline

    \end{tabular}
\caption{Example of the CBC mode of operation evolving over time with $b=2$ blocks, $n=3$ bits and block cipher $E_{K}$.}
\label{fig:ex2}
\end{center}
\end{figure*}

In Fig.~\ref{fig:ex2} we show an illustration of the CBC mode of operation by using Eq.~\eqref{eq:DynamicSystem} for the first two time steps in the evolution of its corresponding CA. Note that text styling has been added to make the effect of the mode of operation tractable, and also, that the same plaintext and $\gamma$ values are used as the ones given in Fig.~\ref{fig:ex1}.
As the mode of operation is applied iteratively, at every consecutive time step $\gamma$ is put equal to $C^{t}_b$ because one has to select a new point in phase space that is the closest to the reference direction, which basically constitutes an orientation of the new initial configuration $C^{t+1}_{0}$ into the direction of $C^{t}_b$. For reasons of uniformity, $\gamma$ is the only parameter that can contain this initial configuration.

Recently, sundry methods have been developed to unravel the dynamical properties of utter discrete dynamical systems such as CA~\cite{Baetens2010,Bagnoli92,courbage2006,langton1990,wolfram84}.
As such, by relying on these for grasping the dynamics of a cryptographic system's corresponding CA, we might be able to gain deeper insight into the dynamics of the former. More specifically, we will show in the remainder of this paper how Lyapunov exponents of CAs may be relied upon for identifying the mode of operation of the underlying cryptographic systems, in the same way as these measures have shown their usefulness for characterizing utter discrete dynamical systems~\cite{Baetens2010,Bagnoli92,Changpin2004,Luque2000}.

\section{A cellular automaton view on cryptographic systems\label{method}}
Suppose that the CA counterpart of a given cryptographic system $\mathscr{G}=\langle \mathbf{C}, \mathbf{P}, M_{E_K} \rangle $ is denoted as $\mathcal{C}$. We can investigate the dynamics of the former in general, and its stability more in particular, by computing its so-called Lyapunov exponent, which quantifies how the dynamical system behaves in the long run if it is evolved from two close initial conditions. Clearly, in our setting, this means that we will assess the sensitivity of the equivalent CA $\mathcal{C}$ to a small perturbation of the plaintext since we put $s(\cdot,0)=\mathbf{P}$. Hence, this should yield insights into the behaviour of the cryptographic system if the plaintext is perturbed.

Taking into account that the smallest possible perturbation in such a two-state setting boils down to flipping one bit of the first block of the plaintext $\mathbf{P}$, $P_1$ constitutes the most influential block (initial condition) to the mode of operation. Besides, it holds that the initial damage vector $h(\cdot,0)=\mathbf{P}\oplus \mathbf{P}^*$,  where $\oplus$ is the mod 2 operator, contains only one non-zero element. In the remainder, a cell $c_i^j$ for which holds that $h(c_i^j,0)=1$ will be referred to as a \textbf{defective cell}. After pinning down a plaintext $\mathbf{P}$ and its perturbed version $\mathbf{P}^*$ that fulfills the latter criterion, we can evolve the equivalent CA $\mathcal{C}$ from both $s(\cdot,0)=\mathbf{P}$ and $s^*(\cdot,0)=\mathbf{P}^*$ for one time step in order to obtain $s(\cdot,1)$ and $s^*(\cdot,1)$. Here, it should be recalled that one time step in the evolution of a CA corresponds to one application of the mode of operation $M_{E_K}$, which involves the update of several blocks. Having updated all the blocks, we can compute the damage vector at the first time step

\begin{equation}
h(c_i^j,1) = \left\{
\begin{array}{l l}
1 & \quad \text{if }  s^{*}(c_i^j,1) \neq s(c_i^j,1),\\
0, & \quad \text{else.}
\end{array} \right. \nonumber \\ \\
\end{equation}

Consequently, the total number of defects at the first time step can be computed as
\begin{equation}
\epsilon_1 = \sum\limits_{j=1}^{b} \sum\limits_{i=1}^{n} h(c^j_i,1)\,.
 \label{eq:damagevector}
\end{equation}

At this point, the reader might think that the damage vector during subsequent time steps should be computed similarly, but this is certainly not the case because one would then neglect the fact that the defects can cancel out each other due to the discrete nature of the CA's state space~\cite{Baetens2010,Bagnoli92}. The discrepancy between the number of defective cells and the number of defects is not yet clear after the first time step because every defective cell at $t=1$ traces back to the same initial defective cell.

\begin{figure}[!htbp]
\begin{center}
\includegraphics[width=0.47\textwidth]{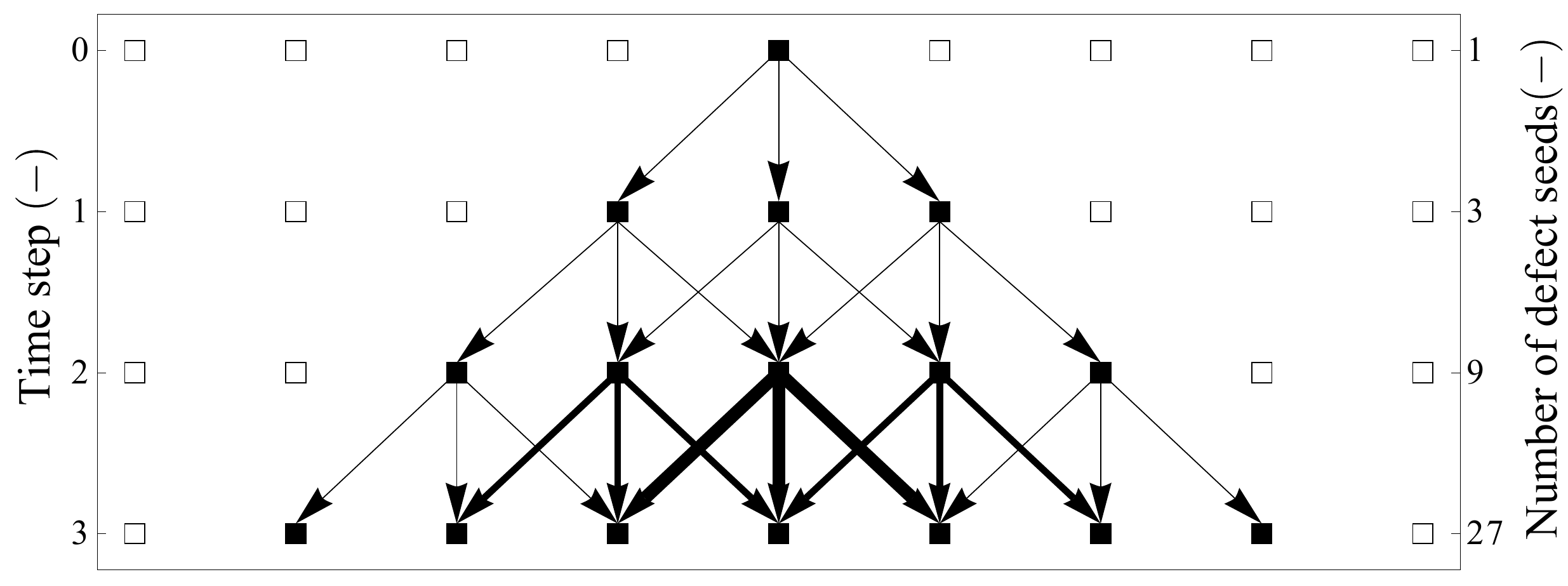}
\caption{Maximal propagation of defects (black) in evolution space of a CA together with all possible pathways along which defects can propagate (arrows).}
\label{propdef}
\end{center}
\end{figure}

However, as soon as the CA is evolved one more time step, a discrepancy emerges between these quantities. This can be understood by explicitly tracking all possible pathways along which defects at $t=1$ may propagate and accumulate during one subsequent time step (see Fig.~\ref{propdef}). It is interesting to have a closer look at how nine defects can arise at the second time step in the evolution of the CA notwithstanding there are only five defective cells, which should be contributed to the fact that several of them may enclose multiple defects due to the existence of several pathways along which defects can propagate.

Taking into consideration this reasoning, the correct number of defects at the $t$-th time step, denoted $\epsilon_t$, should be computed in accordance with the following six-step procedure.

\begin{enumerate}
\item Let the cryptographic system evolve for one time step, \textit{i.e.,} $\mathbf{C}^1 = M_{E_K}(\mathbf{P})$, and analogously for its perturbed version $\mathbf{C}^{1*} = M_{E_K}(\mathbf{P}^*)$, both using the same initialization vector $\gamma$.

\item Calculate the damage vector $h_1$ given by Eq.~\eqref{eq:damagevector} and set the initialization vector as the last block calculated for both versions, \textit{i.e.,} $\gamma= C^1_b$ and $\gamma ^*= C^{1}_b$, as explained in Section \ref{sec:analogies}.

\item For every $c_i^j$ for which $h(c_i^j, 1)=1$, create a replica $R_{i}^{j}$ such that $R_{i}^{j}(c_i^j,1) = s^{*}(c_i^j,1) = \overline{s}(c_i^j,1)$, where $\overline{s}(c_i^j,1)$ is the Boolean complement of $s(c_i^j,1)$, and $R_{i}^{j}(c_q^j,1) = s(c_q^j,1)$, for every $c_{q}^j\neq c_{i}^j$. Use the set $A_1$ to store these replicas.

\item Let the cryptographic system evolve one more time step, \textit{i.e.,} compute $s(\cdot,2)$ and $R(\cdot,2)$, which boils down to evaluating $M_{E_K}(R_{i}^{j})$ with the same $\gamma^*$ for all replicas $R\in A_1$.

\item Calculate the total number of defects at the second time step as follows:
    \begin{equation}
        \epsilon_2 = \sum\limits_{R_{i}^{j}\in A_1}\sum\limits_{j=1}^{b}\sum\limits_{i=1}^{n} h(c_i^j,2)
    \end{equation}
and set the initialization vector as the last block calculated for both versions, \textit{i.e.,} $\gamma= C^2_b$ and $\gamma ^*= C^2_b$.

\item For every $c_i^j$ and $R \in A_{1}$ for which $R(c_i^j,2)\neq s(c_i^j,2)$, create a replica $R_{i}^{j}$ such that $R_{i}^{j}(c_i^j,2) =  \overline{s}(c_i^j,2)$ and $R_{i}^{j}(c_q^j,2) = s(c_q^j,2)$ for every $c_{q}^j\neq c_{i}^j$. Use a multiset $A_2$ to store these replicas.

\item Repeat steps (4)--(6) in every subsequent time step $t+1$ in order to assemble $h(\cdot, t+1)$ and $A_{t+1}$.
\end{enumerate}

After computing the number of defects $\epsilon_t$ at every consecutive time step $t$, the rate of divergence/convergence of initially close phase space trajectories $\lambda(t)$ of a CA $\mathcal{C}$ can be obtained from:
\begin{equation}
\lambda(t) = \frac{1}{t}\log \left( \dfrac{\epsilon_t}{\epsilon_0}\right)\,,
\end{equation}
with its limit value
\begin{equation}
\lambda=\lim_{t\to\infty}\lambda(t)\,,
\end{equation}
generally referred to as the maximum Lyapunov exponent (MLE) of $\mathcal{C}$. In the framework of cryptographic systems, $\lambda(t)$ quantifies how ciphertexts, which are obtained by iteratively encrypting two close plaintexts $\mathbf{P}$ and $\mathbf{P}^*$, behave (converge/diverge) as the number of time steps grows.
As indicated in papers on the LE of CAs, one can derive a theoretical upper bound on these LEs, which has shown to depend on the number of neighbours~\cite{Baetens2010,Bagnoli92}. Calling to mind Shannon's idea of diffusion, which is related to the avalanche effect and states that a slight change of the plaintext gives in worst case rise an entirely different ciphertext, we are able to derive a theoretical upper bound on the MLE of cryptographic systems. For instance, consider the plaintext $\mathbf{P}=10101101$ and its perturbed version $\mathbf{P}^*=10101100$, as well as their encryption $E_K(\mathbf{P})= 10101000$ and $E_K(\mathbf{P}^*)=01010111$, respectively. In the worst case scenario, the number of differences may be at most 8, which is the length of the plaintext $N$. This means that a defect in $c^j_i$ at the $t$-th time step can at most propagate to $N$ bits at the subsequent time step. Therefore, a mean-field approximation of the upper bound $\lambda_m$ for the cryptographic systems becomes

\begin{equation}
\lambda_m= \frac{1}{t}\log(\lambda^t)= \log(N)= \log(bn).
\label{eq:upperbound}
\end{equation}

\noindent Obviously, the higher the number of blocks $b$, the higher the upper bound becomes.

\begin{SCfigure*}
\includegraphics[width=0.72\textwidth,height=0.42\textwidth]{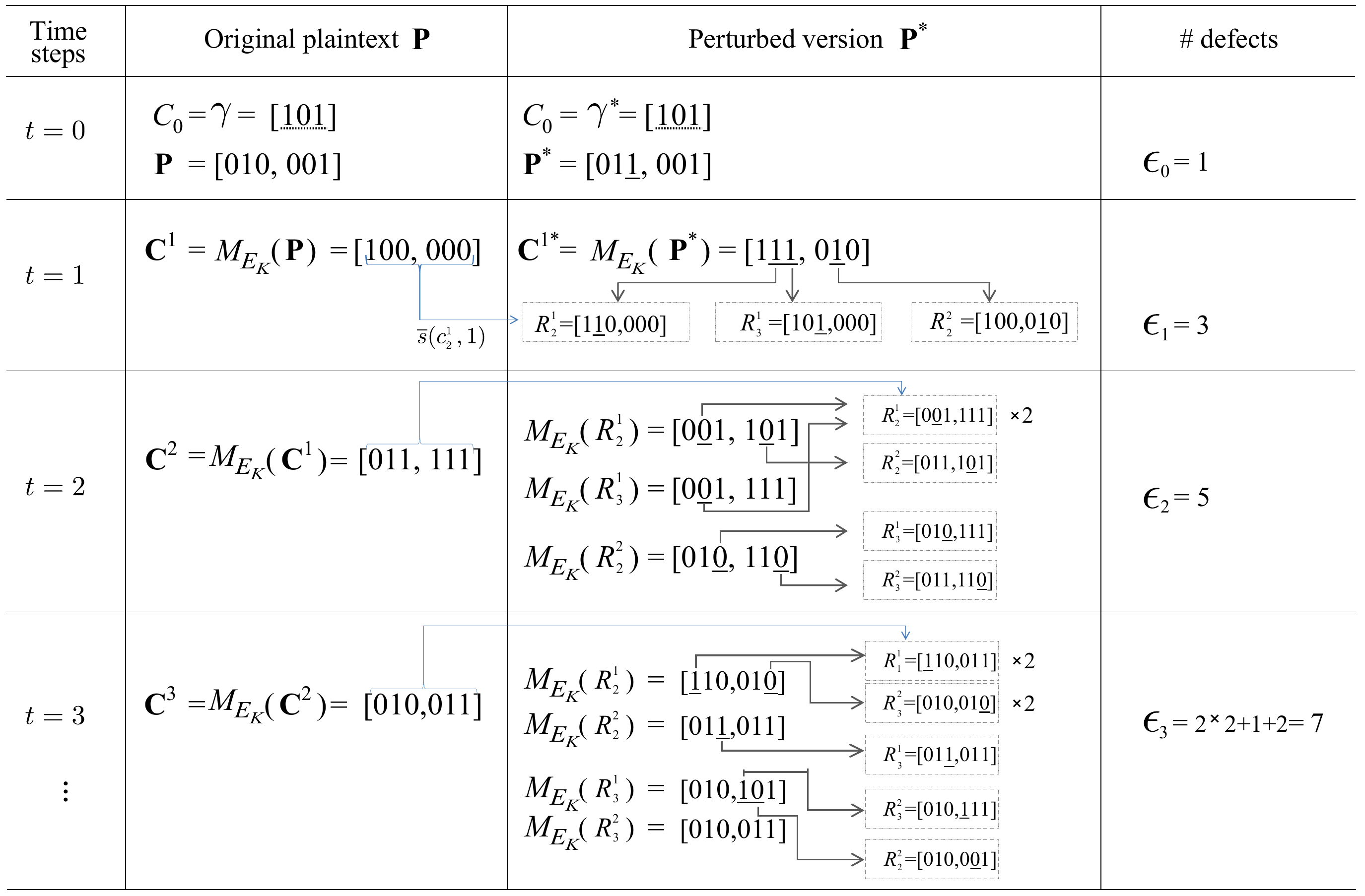}
\caption{Illustration of the calculation of the LE for a generic mode of operation and block cipher with $b=2$ blocks and $n=3$ bits.}
\label{fig:ex3}
\end{SCfigure*}

Before turning to the experimental section of this paper, we illustrate in Fig.~\ref{fig:ex3} the procedure by which the LE of a mode of operation $M_{E_K}$ can be assessed. Note that we employed the same plaintext as in Fig.~\ref{fig:ex2}. By moving along this table's rows, we see the plaintexts that are encrypted by repeatedly applying the mode of operation $M_{E_K}$. Note that some replicas are repeated as soon as the system is evolved for one time step, which is in agreement with the findings reported in Fig.~\ref{propdef}.

\sidecaptionvpos{figure}{c}

In order to avoid having to keep track of all pathways individually, one can optimize the algorithm by just counting every unique replica once and keeping track of its multiplicity. For example, at $t=2$ the replica $R^1_2$ would be repeated twice, then at the subsequent time step this replica has two cell defects, which means four pathways in total at $t=3$. As such, by tracking the multiplicity of every defect rather than every defect itself, the efficiency of the algorithm increases substantially.

If the cryptographic system is sensitive to the plaintext from which it is evolved, the number of defects $\epsilon_t$ increases exponentially during its evolution, such that $\epsilon_t>\epsilon_0$, and consequently, $\lambda>0$. On the other hand, the system should be referred to as insensitive to the underlying plaintext if the initial defect vanishes as the number of time steps increases. Such a situation implies that $\lambda=-\infty$ owing to the discrete nature of a CA's state space. Since cryptographic systems are designed in such a way that small changes in the plaintext give rise to substantially differing ciphertexts, it should not come as a surprise if $\lambda$ would be positive for all the modes of operation at stake in this paper. Still, the numerically assessed values of $\lambda(t)$ might differ significantly among different block ciphers and/or underlying modes of operation, such that they may still be used to discriminate between different modes of operation. This usage will be illustrated in the following section.

\section{Experimental results and discussion\label{res}}

To demonstrate the effectivity of the above methodology, we considered two groups of block ciphers on the basis of their block length $n$. The first group encloses the 64-block ciphers, namely DES~\cite{DES}, IDEA~\cite{IDEA}, TEA~\cite{TEA} and XTEA~\cite{XTEA}, which all use a $k=64$ key except for DES that uses a key of length of $k=56$ bits. The second group of block ciphers is composed of 128-block ciphers with key of $k=128$ bits: AES~\cite{AES2001}, RC6~\cite{RC61998}, Twofish~\cite{Twofish}, Serpent~\cite{Serpent}, Seed~\cite{SEED} and Camellia~\cite{Camellia}.

These ten block ciphers were adapted in such a way that they took the plaintext $\mathbf{P}$, the key $K$ and the initialization vector $\gamma$ as input variables. Moreover, these block ciphers were implemented with a predefined number of \textit{rounds}, each of which consist of several inner steps in the course of an encryption process and they depend on the specific algorithm~\cite{BookCrypto}. We considered eight rounds for IDEA, eighteen for DES, and thirty-two rounds for both TEA and XTEA. Further, ten rounds for AES, sixteen for  both Twofish and Seed, eighteen for Camellia and thirty-two for Serpent. Finally, we implemented the modes of operation in accordance with the formalism given by Eqs.~\eqref{ECBmath}--\eqref{OFBmath} for ECB, CBC and OFB, and in Table~\ref{modestab} for CFB, CTR and PCBC.

\subsection{The dataset}

To ensure the representativity of the computed Lyapunov exponents, we computed the average $\lambda$-values that were obtained when the CAs were evolved from different plaintexts. Hence, the $\lambda$-values reported in the remainder represent averages calculated over an ensemble \linebreak $E=\left\{_{e}\mathbf{P}\mid e=1,\ldots, 200 \right\}$ of 200 randomly generated plaintexts $_{e}\mathbf{P}$ and their perturbed versions $_{e}\mathbf{P}^*$, which were obtained by flipping only one bit in the first block of $_{e}\mathbf{P}$.

Since such an ensemble was constructed for each of the concerned cryptographic systems, by mutually combining the underlying block cipher and the mode of operation, we considered a total of 60 combinations, \textit{i.e.,} ten block ciphers and six modes of operation. Consequently, a total of 12000 plaintexts were generated randomly.

Similarly, the key $K$ and the initialization vector $\gamma$ were generated randomly in order to avoid both key repetitions and weak keys, which is a recommendation in cryptography~\cite{NIST-weak-Key}. In the remainder, we refer to the assembly of the 200 randomly generated plaintexts, keys and initialization vectors as the cryptographic dataset per cipher and mode of operation.
Given that the block ciphers encrypt a plaintext by splitting it into a number of blocks and subsequently transcribing every block, the plaintexts were generated in such a way that they were composed of the same number of $b$ blocks, irrespective of the type of block cipher. Note that this implies a plaintext containing $N=b\times 64$ bits in case of the 64-block ciphers and $N=b\times 128$ bits in case of the 128-block ciphers. Thus, in order to compare $\lambda$-values of 64- and 128-block ciphers, we normalized the numerically obtained $\lambda$ with respect to $\lambda_m$ obtained from Eq.~\eqref{eq:upperbound},

\subsection{Lyapunov exponents for the modes of operation}

We assessed $\lambda(t)$ for the CA counterpart of each of the concerned cryptographic systems. For that purpose, the equivalent CAs were evolved for $t=200$ time steps, which was sufficiently long because $\lambda(t)$ showed convergence in such a way that $\left|\lambda(t)-\lambda(t+1)\right|<1.19\times 10^{-4}$, being the maximum discrepancy between $\lambda(t)$ and $\lambda(t+1)$ at the end of the simulation.

\begin{figure}[!htbp]
\begin{center}
\includegraphics[width=0.48\textwidth,height=0.2\textwidth]{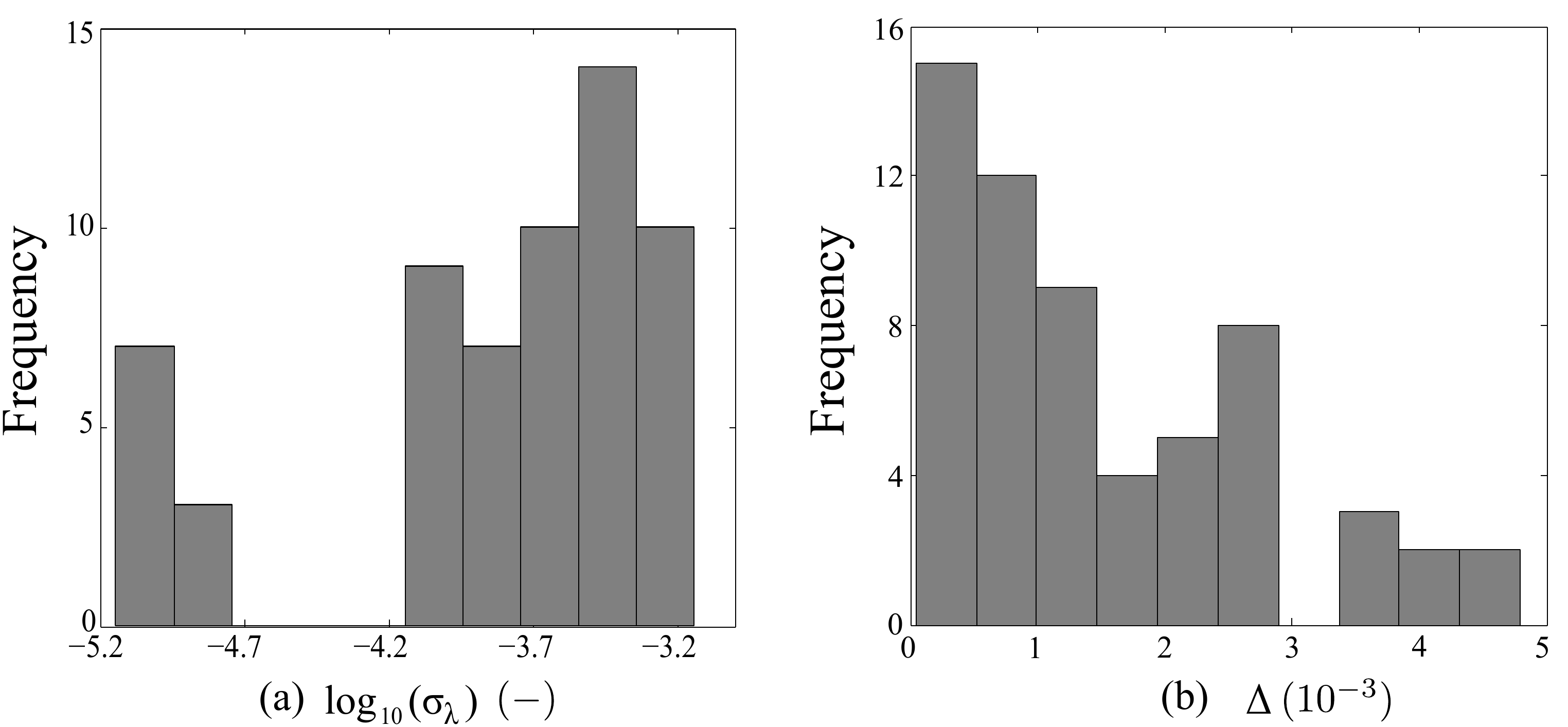}
\caption{(a) Frequency distribution of the standard deviation $\sigma_{\lambda}$ (base 10 logarithm) of $\lambda_{200}$ calculated over the 60 combinations of cryptographic systems. (b) Frequency distribution of
  $\Delta = \max(\{_{e}\mathbf{P}\mid e=1,\ldots,200\}) - \min(\{_{e}\mathbf{P}\mid e=1,\ldots,200\})$.}
\label{histoDeltaLog}
\end{center}
\end{figure}

Furthermore the consistency of the $\lambda$-values across the members of the different ensembles is also demonstrated in Fig.~\ref{histoDeltaLog} (a), which depicts the frequency distribution of the standard deviation $\sigma_{\lambda}$ (base 10 logarithm) of  $\lambda_{200}$ calculated for the 60 cryptographic systems at stake. This histogram shows the similarity of $\lambda$ across the different plaintexts, as the standard deviation that comes along with the average MLE for the different cryptographic systems is obviously small.

Fig.~\ref{histoDeltaLog} (b) shows a histogram of the discrepancies between the maximum and minimum observed MLE among the members of the ensemble $E$, denoted by $\Delta = \max(\left\{_{e}\mathbf{P}\mid e=1,\ldots,200\right\}) - \min(\left\{_{e}\mathbf{P}\mid e=1,\ldots,200\right\})$.

The $\Delta$-values obtained for the investigated datasets lie between $ 3.99\times 10^{-5}$  and $4.79\times 10^{-3}$. The maximum and minimum $\Delta$ correspond, respectively, to the pairs of modes of operation with underlying block cipher: IDEA-OFB and Serpent-CFB. All together, these small discrepancies demonstrate once more that the $\lambda$-values are highly consistent across the different plaintexts within the ensemble $E$, such that we may draw conclusions on their average values.

\begin{figure}[!ht]
\center
\begin{tabular}{c}
\subfigure[IDEA]
{\includegraphics[width=0.37\textwidth,height=0.28\textwidth]{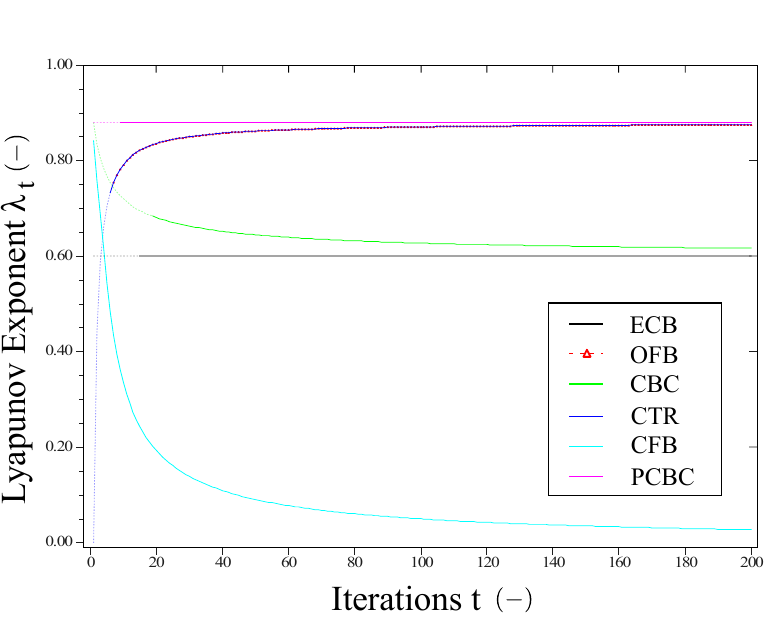} \label{fig:IDEA}}
\\
\subfigure[AES]
{\includegraphics[width=0.37\textwidth,height=0.28\textwidth]{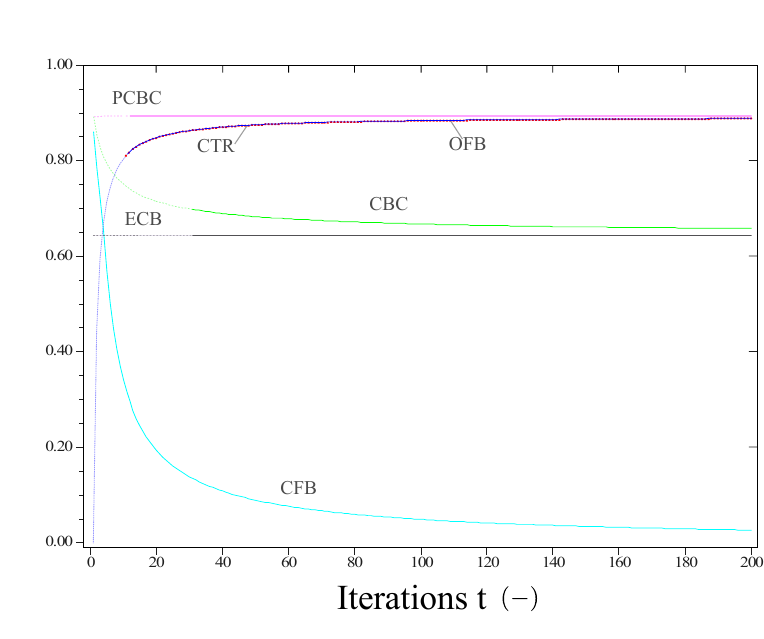} \label{fig:AES}}
\end{tabular}
\caption{Lyapunov exponent versus the number of time steps for different modes of operation, being ECB, CBC, OFB, CFB, CTR and PCBC, using the (a) IDEA 64-block cipher and (b) AES 128-block cipher. The curves represent the average LE $\lambda$ over 200 initial plaintexts during $t=200$ times steps, which are normalized with respect to $\lambda_m$ obtained from Eq.~\eqref{eq:upperbound}.}
\label{fig:IDEAAES}
\end{figure}

\begin{SCfigure*}
\includegraphics[width=0.65\textwidth, height=1.05\textwidth]{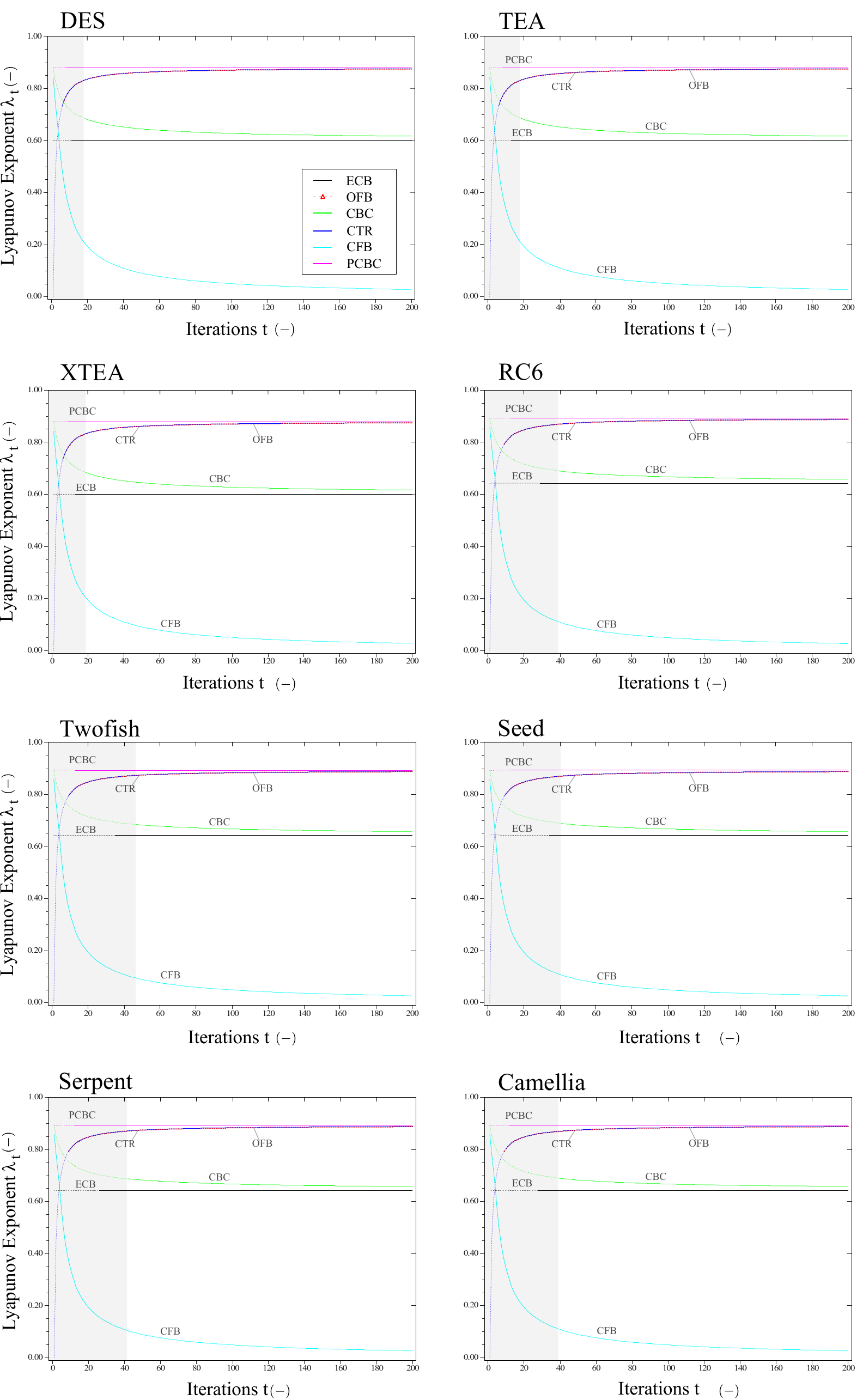}
\caption{Lyapunov exponent versus the number of time steps for different modes of operation, being ECB, CBC, OFB, CFB, CTR and PCBC using 64-block ciphers: DES, TEA and XTEA, and 128-block ciphers: RC6, Twofish, Seed, Serpent and Camellia with $b=5$ blocks and $n=64$ bits. The curves represent the average LE $\lambda$ over 200 initial plaintexts during $t=200$ time steps, which are normalized with respect to $\lambda_m$ obtained from Eq.~\eqref{eq:upperbound}}
\label{fig:others}
\end{SCfigure*}

For simplicity, we mainly restrict our attention to IDEA and AES as representative members of the families of 64- and 128-block ciphers (see Fig.~\ref{fig:IDEAAES}), respectively, since similar results were obtained for the other block ciphers in the same families (see Fig.~\ref{fig:others}). These figures depict the LE curves for all the modes of operation versus the number of time steps---note that the standard deviation of each curve is not shown since they are very small. Taking into account Eq.~\eqref{eq:upperbound}, the upper bound for the 64-block ciphers is $\lambda_m=\log(N)=\log(320)\approx 5.76832$, while it is $\lambda_m=\log(640)\approx 6.46147$ for the 128-block ciphers were employed to normalize the $\lambda$-values, respectively.

\sidecaptionvpos{figure}{c}

\begin{table*}[hbt]
  \caption{MLE $\lambda_{200}$ and standard deviation ($\times 10^{-3}$) for the (a) 64- and (b) 128-block ciphers.}
  \label{tab:MLE2}
  \begin{center}
    \subtable[\label{tab:sub1A1}]{
    \begin{tabular}{|c|c|c|c|c|}
\hline
    $\lambda_{200} $ & IDEA & DES & TEA & XTEA  \\
\hline
    ECB          & 3.46565 $\pm$ 1.35 & 3.46570 $\pm$ 1.27 & 3.46554 $\pm$ 1.23 & 3.46564 $\pm$ 1.25 \\
    OFB          & 5.04796 $\pm$ 4.00 & 5.04800 $\pm$ 3.89 & 5.04817 $\pm$ 3.87 & 5.04816 $\pm$ 3.68 \\
    CBC          & 3.55597 $\pm$ 0.96 & 3.55598 $\pm$ 0.93 & 3.55596 $\pm$ 0.81 & 3.55598 $\pm$ 0.88 \\
    CTR          & 5.04817 $\pm$ 3.55 & 5.04818 $\pm$ 3.81 & 5.04853 $\pm$ 4.14 & 5.04787 $\pm$ 4.15 \\
    CFB          & 0.15925 $\pm$ 0.07 & 0.15925 $\pm$ 0.07 & 0.15926 $\pm$ 0.06 & 0.15925 $\pm$ 0.07 \\
    PCBC         & 5.07496 $\pm$ 1.91 & 5.07465 $\pm$ 2.01 & 5.07467 $\pm$ 1.91 & 5.07451 $\pm$ 1.87 \\

    \hline
\end{tabular}
}\\

\subtable[\label{tab:sub1B}]{
\begin{tabular}{|c|c|c|c|c|c|c|}
    \hline
    $\lambda_{200}$ & AES & RC6 & Twofish & Seed & Serpent & Camellia\\
    \hline
    ECB & 4.15879 $\pm$ 0.67 & 4.15885 $\pm$ 0.75 & 4.15876 $\pm$ 0.68 & 4.15887 $\pm$ 0.7 & 4.15887 $\pm$ 0.75 & 4.15884 $\pm$ 0.76 \\
    OFB & 5.73875 $\pm$ 2.82 & 5.73879 $\pm$ 2.96 & 5.73869 $\pm$ 2.86 & 5.73860 $\pm$ 2.84 & 5.73840 $\pm$ 2.82 & 5.73885 $\pm$ 3.02 \\
    CBC & 4.24921 $\pm$ 0.52 & 4.24913 $\pm$ 0.56 & 4.24920 $\pm$ 0.56 & 4.24915 $\pm$ 0.51 & 4.24919 $\pm$ 0.56 & 4.24920 $\pm$ 0.55 \\
    CTR & 5.73891 $\pm$ 2.83 & 5.73876 $\pm$ 2.70 & 5.73882 $\pm$ 2.79 & 5.73904 $\pm$ 2.78 & 5.73866 $\pm$ 2.74 & 5.73871 $\pm$ 2.81 \\
    CFB & 0.17311 $\pm$ 0.05 & 0.17311 $\pm$ 0.05 & 0.17311 $\pm$ 0.05 & 0.17311 $\pm$ 0.05 & 0.17311 $\pm$ 0.05 & 0.17311 $\pm$ 0.05 \\
    PCBC & 5.76812 $\pm$ 1.31 & 5.76815 $\pm$ 1.27 & 5.76792 $\pm$ 1.37 & 5.76816 $\pm$ 1.40 & 5.76792 $\pm$ 1.38 & 5.76806 $\pm$ 1.30 \\
    \hline
\end{tabular}
}
\end{center}
\end{table*}

From both figures, it is quite easy to discriminate between the modes of operation, except for the overlapping curves of CTR and OFB. It is interesting to have a closer look at how their corresponding LE increases almost exponentially during the first few time steps. This behaviour can be explained by reconsidering the equations in Table~\ref{modestab}. Indeed, CTR uses an inner pseudo-random number generator and OFB uses a new initialization vector at every time step, such that the number of defective cells will not only increase due to discrepancies naturally emerging between the plaintexts, but also due to the additional defects introduced through these random processes. In contrast, CFB displays the opposite effect, \textit{i.e.,} its curve decays exponentially.

Further, the Lyapunov exponents of ECB and PCBC are constant, since the number of defective cells grows proportionally through time. For instance, this behaviour can be explained by considering that ECB attains $\lambda_{200}=3.46565 \pm 1.35\times 10^{-3}$ with the IDEA block cipher, which means that the number of defects $\epsilon_t = e^{(\lambda_{200})}= 31.99728 \pm 4.32\times 10^{-2}$ equals approximately $n/2$, being almost 50\% of the block length. The same behaviour occurs for the AES block cipher, where $e^{(\lambda_{200})}= 63.994055 \pm 4.29\times 10^{-2}$.

The original values of the exponents and standard deviation found at the end of the simulation are summarized in Table~\ref{tab:MLE2}. Furthermore, we can discriminate between modes of operation in families of block ciphers not only by visual inspection of the graphs, but also by means of statistical tests to demonstrate it. Paired t-tests revealed that there are statistically significant differences between ECB, CBC, CFB and PCBC mutually at the 5\% significance level. However, no statistically significant difference was found between CTR and OFB.
These observations were found in both families of 64- and 128-block ciphers, which also allows the proposed method to discriminate between modes of operation of different families, for instance, IDEA-CTR can be distinguished from AES-CTR.

Moreover, by means of the washer method~\cite{Washer}, we analyzed statistically whether a curve (for one of the ensemble's members) for a given mode falls within the envelope of curves obtained for the entire ensemble for another mode of operation. This procedure was repeated for all the curves within the given mode of operation at the 5\% significance level, which indicated that at most 6.4\% and 11.1\% of the observations of the 64- and 128-block ciphers constituted outliers. These outliers exist at the beginning of the simulation. We can also notice that the CFB curve behaves completely different from the other ones (non overlapping and not too close), which means that the CFB falls outside of the range of the LE for the other modes of operation.

\subsection{Analysis of the LE of cryptographic systems}

In this section, we will examine and discuss the preceding results in more detail.

\paragraph{\textbf{Initial conditions $\mathbf{P}$ and $\mathbf{P^*}$}}

    Recalling that an assessment of the Lyapunov exponent involves tracking the CA evolution from two initial configurations for which it holds that $h(\cdot,0)=\mathbf{P}\oplus \mathbf{P}^*$ and given the fact that we are dealing with Booleans, the smallest possible perturbation in such a setting implies flipping the right-most bit of the plaintext. However, we observed that the position of the flipped bit does not affect the numerical value of the Lyapunov exponent, which can be understood by recalling the theoretical upper bound on the Lyapunov exponent of cryptographic systems. Indeed, the encryption of the plaintext with a bit flipped at an arbitrary $x$ position at the first plaintext block ($1\leq x\leq n$) may, in the worst case, affect all the bits of the ciphertext.

\paragraph{\textbf{Cryptographic signatures}}

    From Figs.~\ref{fig:IDEAAES}--\ref{fig:others} it should be noticed that the normalized $\lambda$-values do not attain 1, because none of the cryptographic systems at stake is attaining the worst case scenario mentioned in Section~\ref{method}. This can be explained by the fact that this is an ideal and desirable ``scenario'' for cryptographic systems and that the block ciphers at stake have their limitations. Furthermore, we found an unexpected phenomenon with regard to the families of 64- and 128-block ciphers. A closer look at the curves tied up with these families shows that, in fact, we can also differentiate between block cipher families.

\paragraph{\textbf{Number of blocks}}

\begin{figure*}[!t]
    \begin{center}
       \includegraphics[width=0.73\textwidth,height=0.83\textwidth]{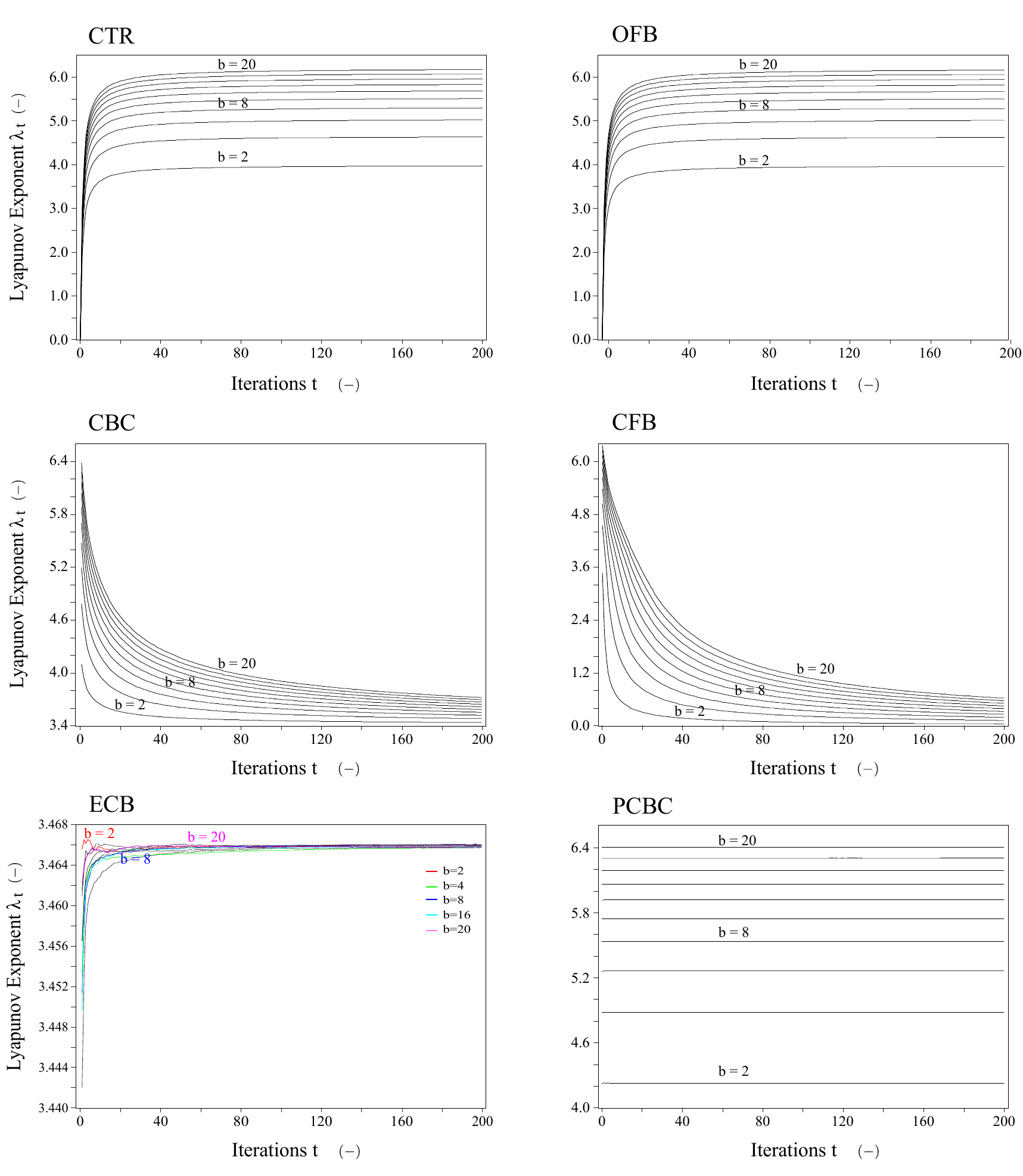}
       \caption{Comparison of the LE for different modes of operation with $2,4,8,\ldots,20$ blocks using IDEA. Each plot contains curves which represent the average LE $\lambda$ over 200 initial plaintexts during $t=200$ time steps of the IDEA 64-block cipher.}
      \label{fig:compareModes}
    \end{center}
    \end{figure*}

     In order to investigate the importance of the number of blocks $b$, we computed the Lyapunov exponents for the same cryptographic systems but with $2,4,8,\ldots,20$ blocks. In Fig.~\ref{fig:compareModes} one can see the LE curves for each mode of operation using the IDEA block cipher as a function of $b$. From this figure it is clear that each curve is somehow lifted upwards with the number of blocks $b$.

     From Eq.~\eqref{eq:upperbound}, it is expected to get higher $\lambda$-values when the length of the plaintext increases. In fact, this occurs for all the modes of operation at stake, except for ECB. In particular, this indicates that irrespective of the number of blocks $b$ of the plaintext, the ECB's number of defects will attain approximately $n/2$, which demonstrates that the first block is the only one affected by the initial conditions, while the other blocks do not spread defects. Furthermore, we obtained the same results with both groups of 64- and 128-block ciphers and the six modes of operation.

\paragraph{\textbf{Empirical MLE}}

    In Section~\ref{method} we provided some mathematical foundations to obtain the theoretical upper bound $\lambda_m$ for cryptographic systems. However, the lack of an analytical upper bound for a specific mode of operation is an important limitation remaining. Perhaps, this limitation may seem contradictious to cryptography aims, because it indicates to find an analytical way to obtain the essence of a cryptographic system that is designed to avoid this gap.

\begin{figure*}[ht!]
    \begin{center}
    \includegraphics[width=0.85\textwidth,height=0.9\textwidth]{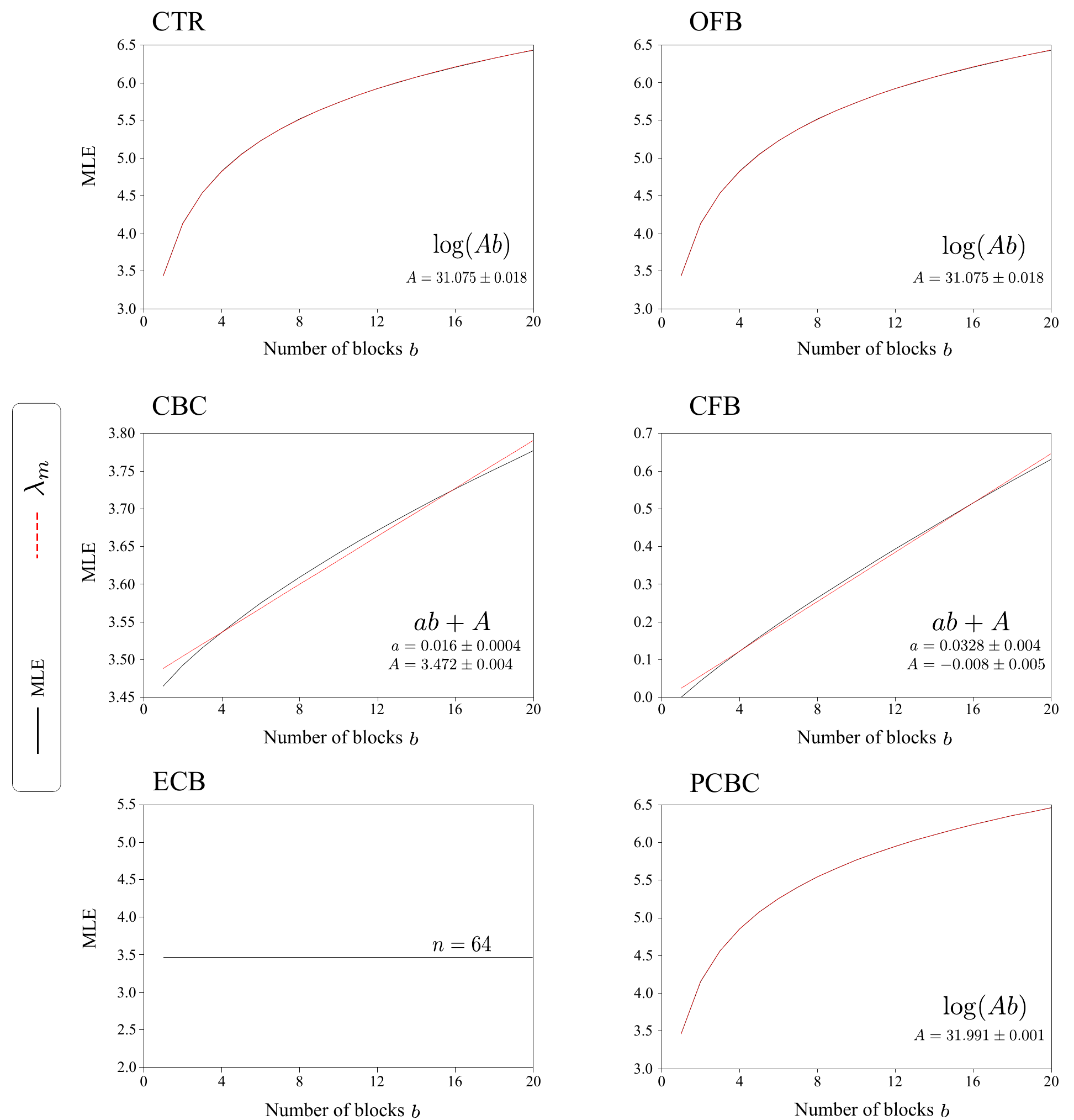}
    \caption{An empirical analysis of the upper bound for the modes of operation based on IDEA $\lambda_{200}$ with different number of blocks $(R^2=0.99)$.}
    \label{fig:upper64}
    \end{center}
    \end{figure*}

\begin{figure*}[!ht]
    \begin{center}
    \includegraphics[width=0.85\textwidth,height=0.9\textwidth]{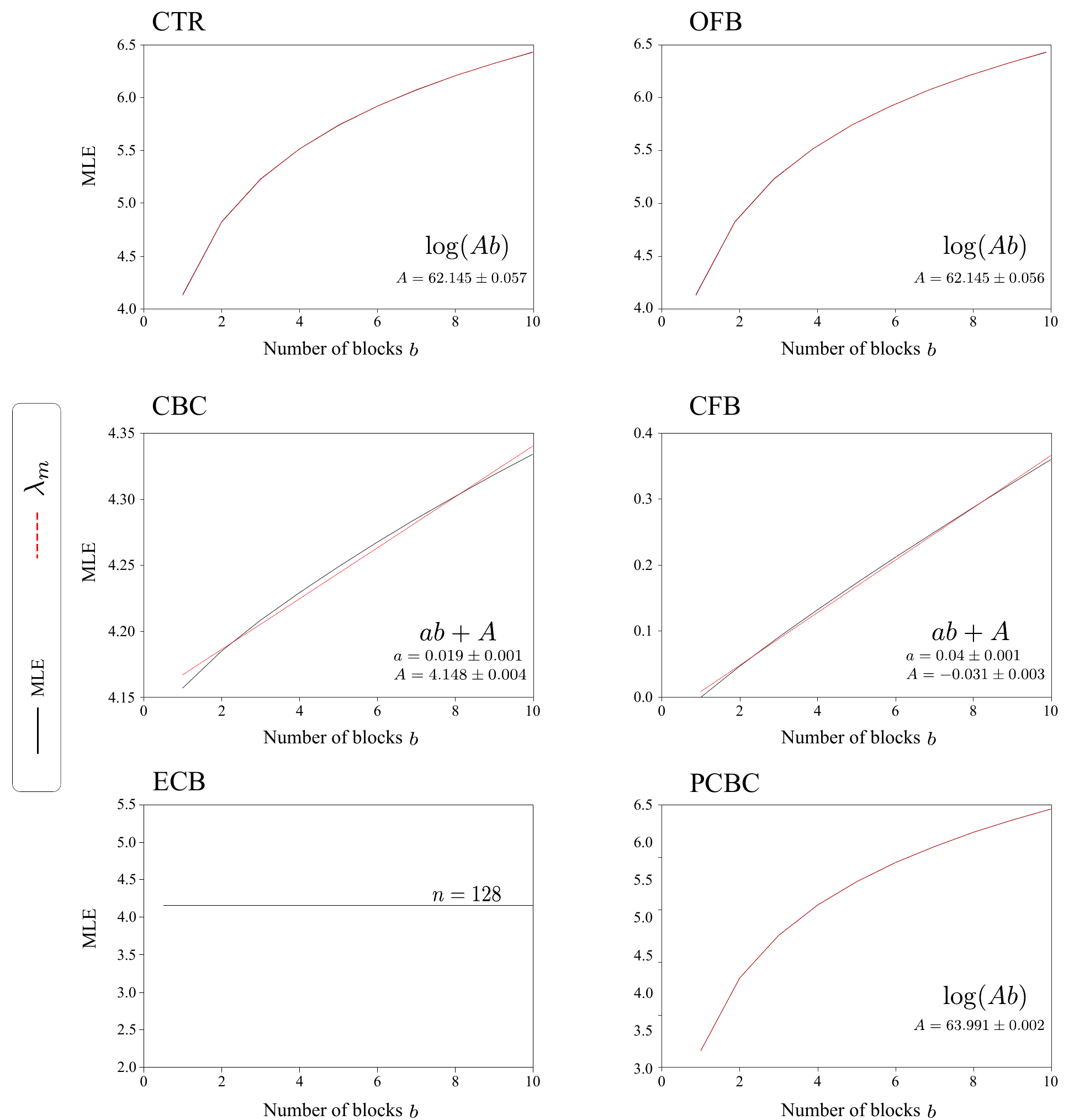}
    \caption{An empirical analysis of the upper bound for the modes of operation based on AES $\lambda_{200}$ with different number of blocks $(R^2=0.99)$.}
    \label{fig:upper128}
    \end{center}
    \end{figure*}

    Here, we opted for an alternative approach, \textit{i.e.,} an empirical estimation of the upper bound for the modes of operation as a function of the number of blocks, based on a multiple regression analysis, to gain a deeper insight into the behaviour of the LE curves. In Fig.~\ref{fig:upper64} we show the MLE $\lambda$ obtained as a function of $b$ for the IDEA and the corresponding regression curves (see also Fig.~\ref{fig:upper128} for the AES block cipher). The coefficients of determination also demonstrate the strong influence by the number of blocks $b$.

\newpage\leavevmode\thispagestyle{empty}\newpage

\newpage\leavevmode\thispagestyle{empty}\newpage

\paragraph{\textbf{Computational considerations}}

Although the algorithm has been enhanced with different strategies to avoid tracking each individual defect, somehow the computational cost of the proposed methodology is very high, since the computational time dramatically increases as the number of blocks of the plaintext $\mathbf{P}$ grows.

\section{Conclusions\label{conclu}}

There is a strong relationship between cryptographic systems and discrete dynamical systems. In this work we have outlined an approach to envisage a cryptographic system as an equivalent one-dimensional CA in order to assess its stability characteristics by computing the Lyapunov exponent of the cryptographic system. The proposed method was capable of distinguishing six cryptographic modes of operation, namely ECB, CBC, OFB, CFB, CTR and PCBC using two families of block ciphers DES, IDEA, TEA and XTEA of 64 bits, as well as AES, RC6, Twofish, Seed, Serpent and Camellia of 128 bits. Moreover, the proposed method is also capable of distinguishing between the two families of 64- and 128-block ciphers. The results showed that the Lyapunov exponent evolution pattern is maintained for each mode of operation and this is independent of the block cipher used.

We also provided a mathematical basis to obtain the theoretical upper bound $\lambda_m$ for the cryptographic systems. However, further work is required to theoretically analyze the upper bound on the LE for each of the modes of operation. Here we only used an empirical assessment to fit the curves. Finally, our results suggest that even modern and contemporary algorithms yield patterns that should be explored. Thus, our theoretical framework may offer a novel alternative to explore the weakness of these cryptographic systems and may ultimately lead to a classification of these systems according to their strength.

\section*{Acknowledgements}
J. M. acknowledges support from FAPESP (The State of S\~{a}o Paulo Research Foundation) (2011/05461-0). O. M. B. acknowledges support from CNPq (Grant \#307797/2014-7 and \#484312/2013-8) and FAPESP (Grant \# 2011/01523-1). This work has benefited from the cooperation agreement that has been established between Universidade de S\~{a}o Paulo and Ghent University.

\end{document}